\documentclass{article}
\usepackage{waspaa23,amsmath,graphicx,url,times}
\usepackage{color}
\usepackage{enumitem}
\usepackage{booktabs}
\usepackage{multirow}
\usepackage{hhline}
\usepackage[T4,T1]{fontenc}
\makeatletter
\newcommand{\do@openE}[1]{%
  \mbox{\fontsize{#1}\z@\usefont{T4}{cmr}{m}{n}\symbol{130}}%
}
\newcommand{\openE}{\mathord{\mathchoice
  {\do@openE\tf@size}
  {\do@openE\tf@size}
  {\do@openE\sf@size}
  {\do@openE\ssf@size}
}}
\usepackage{ifthen}
\title{Sound source distance estimation in diverse and \\ dynamic acoustic conditions}

\name{Saksham Singh Kushwaha$^{1,2}$,
        Iran R. Roman$^{2}\sthanks{corresponding author email: roman@nyu.edu}$,
        Magdalena Fuentes$^{2,3}$,
        Juan Pablo Bello$^{2}$
      }
\address{
        $^1$ Courant Institute of Mathematical Sciences, New York University, NY, USA \\
        $^2$ Music and Audio Research Lab, New York University, NY, USA \\    
        $^3$ Integrated Design and Media, New York University, NY, USA
      }

\begin{document}

\ninept
\maketitle

\begin{sloppy}

\begin{abstract}
Localizing a moving sound source in the real world involves determining its direction-of-arrival (DOA) and distance relative to a microphone. Advancements in DOA estimation have been facilitated by data-driven methods optimized with large open-source datasets with microphone array recordings in diverse environments. In contrast, estimating a sound source’s distance remains understudied. Existing approaches assume recordings by non-coincident microphones to use methods that are susceptible to differences in room reverberation. We present a CRNN able to estimate the distance of moving sound sources across multiple datasets featuring diverse rooms, outperforming a recently-published approach. We also characterize our model’s performance as a function of sound source distance and different training losses. This analysis reveals optimal training using a loss that weighs model errors as an inverse function of the sound source true distance. Our study is the first to demonstrate that sound source distance estimation can be performed across diverse acoustic conditions using deep learning.
  
\end{abstract}


\section{Introduction}
\label{sec:intro}
Sound source localization (SSL) \textemdash \ the task of localizing the position of a sound source relative to a microphone \textemdash \ has been an active area of research for decades \cite{wax1983optimum,huang2001real,adavanne2018sound}.
SSL has important downstream applications, including sound source separation \cite{jenrungrot2020cone}, audio-based navigation systems \cite{chen2020soundspaces}, and urban surveillance \cite{valenzise2007scream}. 
SSL can be broken down into two subtasks: direction-of-arrival (DOA) estimation, which approximates sound directivity in terms of azimuth and elevation angles, and distance estimation, which approximates the separation between the sound source and the microphone.

Recent developments have focused in DOA estimation. 
This includes large open-source datasets with DOA annotations for moving sound events in real \cite{evers2020locata,politis_archontis_2023_7709052} and simulated \cite{Adavanne2019_DCASE,politis2021dataset} acoustic conditions. 
Using these datasets, researchers have developed models able to simultaneously carry out DOA estimation and classification (i.e. speech vs music vs engine, etc.) \cite{nguyen2022salsa,shimada2022multi}.
In contrast, distance estimation remains understudied, partly because it is considered to be more difficult \cite{grumiaux2022survey}.
While recent DOA approaches, notably those developed in the context of the DCASE challenge \cite{politis_archontis_2023_7709052,Adavanne2019_DCASE,politis2021dataset,nguyen2022salsa,shimada2022multi}, output 3D coordinates to localize sources, they assume those sources to be in the unit sphere, effectively only estimating azimuth and elevation (i.e. DOA). To the best of our knowledge, existing distance estimation approaches include signal processing methods that assume a room's $T_{60}$ to contrast sounds reaching a microphone directly versus indirectly \cite{liu2021study,chen2016direct}. 
Data-driven approaches have also been developed, but using small datasets that feature only a handful of rooms \cite{yiwere2017distance,roden2019sound}, or framing the task as classification instead of directly estimating distance \cite{takeda2016sound,yiwere2017distance,roden2019sound,bologni2021acoustic}.

While popular datasets used for DOA estimation lack distance annotations, many have metadata from where this information can be recovered.
In this study we add distance annotations to existing open-source datasets. 
We use these to optimize a convolutional recurrent neural network (CRNN) that estimates the distance of moving sound sources from tetrahedral microphone recordings. 
Ours system is the first of its kind (i.e. using deep learning), being able to carry out the task of distance estimation, and evaluated in diverse acoustic conditions. 
We also analyze the effect of different loss functions to learn the task.
Our model outperforms a recent distance estimation approach \cite{daniel2022echo} evaluated on the open-source LOCATA dataset \cite{evers2020locata}.
Additionally we evaluate our model's performance across other datasets.
In summary, our contributions are:
\begin{enumerate}[nolistsep]
\item Distance annotations for a collection of open-source datasets previously used for DOA estimation.
\item A model able to estimate the distance of sound sources in diverse environments and acoustic conditions.
\item An analysis of model performance resulting from using different loss functions.\footnote{Code and data:$_\texttt{github.com/sakshamsingh1/sound\textunderscore distance\textunderscore estimation}$} 
\end{enumerate}

\section{Related Work}
\label{sec:related}

Sound source distance estimation is straightforward if the onset time $t_o$ and the speed of sound $c$ are known. 
In a microphone recording, the sound would appear at time $t_r>t_o$.
The sound source distance $d$ can be calculated by $d = c \times (t_r-t_o)$.    
In the real world, however, knowing $t_o$ is virtually impossible. 

We focus on sound source distance estimation in enclosed, reverberant environments. 
Early approaches were inspired by human listening. 
Humans use the direct-to-reverberant ratio (DRR) \cite{sheeline1983investigation,zahorik2002direct}, which is the ratio between the signal energy directly reaching the listener and energy from wall reflections. 
The DRR can be applied to multi-channel recordings to carry out sound source distance estimation \cite{liu2021study,chen2016direct}. 
Alternatives include  binaural cues like spectral magnitude difference \cite{georganti2013sound} and signal coherence \cite{vesa2007sound}. 
More recently, data-driven approaches have been proposed, including feedforward neural networks (FNNs) or convolutional neural networks (CNNs) with a classification output to categorize sound source distances into one of $N$ pre-defined distance ranges \cite{yiwere2017distance,takeda2016sound,roden2019sound,bologni2021acoustic}.
These models have been developed and evaluated using synthetic datasets (i.e. using simulated wave propagations) \cite{bologni2021acoustic} or recordings in a handful of rooms with specific microphone and loudspeaker configurations \cite{takeda2016sound,yiwere2017distance,roden2019sound}. 


When it comes to DOA estimation, many studies have used CRNNs\footnote{ 
The survey by Grumiax et al. \cite{grumiaux2022survey} reviews all relevant SSL literature.}
that estimate x, y, z coordinates on an assumed unit sphere (i.e. only estimating azimuth and elevation) \cite{shimada2022multi,nguyen2022salsa,Adavanne2019_DCASE,politis2021dataset}.
These approaches benefit from open-source data, sometimes produced by
generators able to yield large-scales of training data \cite{Adavanne2019_DCASE,politis2021dataset}. 
Generators use real-world multi-channel impulse responses (IR) and noise samples collected in different rooms. Sound scenes can be produced where events can be stationary or move along trajectories traced along neighboring IRs.
Datasets with real recordings in rooms also exist \cite{politis_archontis_2023_7709052,evers2020locata}, and are used to evaluate models in real-world contexts.

Besides DOA estimation, these datasets could also be used to develop distance estimation methods.
For instance, Daniel et al. \cite{daniel2022echo} developed a technique that compares higher-order ambisonics (HOA) channels (4th order; 25 total channels) to find temporal relations between a sound source's wall reflections and infer the delay of the propagating signal.
This representation is called the Generalized Time-domain Velocity Vector (GTVV) \cite{kitic2022generalized}.
While their implementation is not publicly-available, they did evaluate it on LOCATA \cite{evers2020locata}, allowing for future comparison between methods using this dataset as a point of reference.

\section{Methods}


\begin{table}
 \begin{center}
 \begin{tabular}{lccccccc}
   \toprule\toprule
   \textbf{Dataset} & Range & Avg & Ntr & Nts & L & R & M\\ 
   \midrule\midrule
    DCASE  & 1.35-7.15 & 3.34 & 900 & 300 & 60.0 & 9 &Y    \\ \hline
    STARSS     & 0.42-7.02& 1.83 & 87 & 74 & 162.2 &16&Y     \\ \hline
    LOCATA     & 0.50-3.49 & 1.78 & 27 & 5 & 18.9 & 1 &Y     \\ \hline  
    MARCo     & 2.6-12 & 4.01 & 5 & 7 & 78.6 &1&N \\ \hline    
    METU      & 0.3-2.2 & 1.41 & 146 & 98 & 2.0 & 1 &N  \\ \hline  
   \bottomrule
   \end{tabular}
\end{center}
\setlength{\belowcaptionskip}{-15pt}
 \caption{Summary of datasets used in our study. Columns indicate the range of distances (``Range'') and average distance (``Avg'') (both in meters), the number of training (``Ntr'') and test (``Nts'') recordings in each dataset, the average recording duration (``L'') (in seconds), the number of unique rooms featured in the dataset (``R''), and whether sound sources move (``M'') (Y: yes; N: no). DCASE and STARSS are split into training, validation, and test sets, each with a unique set of rooms.}
 \label{datasets}
\end{table}

\subsection{Datasets}


We annotate sound source distances in existing open-source datasets and a data generator featuring single sound events in real, dynamic, and diverse rooms. 
We select datasets that use EigenMike since it has been commonly used for DOA estimation research \cite{adavanne2018sound,grumiaux2022survey,politis_archontis_2023_7709052}. 

We use the open-source data generator\footnote{$_\texttt{github.com/danielkrause/DCASE2022-data-generator}$} by Politis et al. \cite{politis2021dataset}. 
It places sounds in nine unique rooms with predefined trajectories where sounds can appear featuring different power levels. 
We modified its code to annotate the sound source distance, which we inferred via each room's metadata files where the possible trajectories are delineated.
With this generator we create a ``DCASE'' dataset with recordings separated into training and test splits, each using a different set of rooms.
\begin{table}
 \begin{center}
 \begin{tabular}{lll}
   \toprule\toprule
   Acronym & Full name&$\openE$\\ 
   \midrule\midrule
   AE & absolute error & $|y-\hat{y}|$ \\
   \midrule
   SE & squared error & $(y-\hat{y})^2$ \\
      \midrule
   APE & absolute percent error & $\frac{1}{y}|y-\hat{y}|$\\
      \midrule
   SPE & squared percent error & $(\frac{1}{y}(y-\hat{y}))^2$\\
      \midrule
   TAPE & thresholded APE & $\max(\delta,\frac{1}{y}|y-\hat{y}|)$\\
   \bottomrule\bottomrule
   \end{tabular}
\end{center}
\setlength{\belowcaptionskip}{-15pt}
 \caption{Different regressors $\openE$ that we investigate in the loss function. We try TAPE with $\delta=0.01$, $\delta=0.1$, $\delta=0.20$.}
 \label{losses}
\end{table}
We also use four datasets featuring recordings in real-world environments.
We use STARSS (2023 version) \cite{politis_archontis_2023_7709052}, which contains sound source distance annotations by its original authors \cite{politis_archontis_2023_7709052}. 
It features recordings in sixteen unique rooms. 
We only estimate the distance of single sound sources. 
Therefore we masked samples where overlapping sounds are present by replacing them with the room's ambient noise.
For STARSS we used the ``development'' set, which comes with recordings separated into training and test splits.
LOCATA \cite{evers2020locata} features recordings in a single room and contains metadata files encoding the sound source distance.
We split it by using all ``Task 1'' and ``Task 5'' files for training, and ``Task 3 evaluation'' for testing (consistent with \cite{daniel2022echo} to compare performance)\footnote{Other ``tasks'' in LOCATA feature overlapping sounds.}.
The 3D-MARCo dataset \cite{lee2019open} contains recordings of musical performances inside a reverberant church. 
We consulted the dataset's documentation and original authors to determine the precise sound source distance.
We used the ``single sources'' recordings for testing and the rest for training. 
We excluded the ``trio'' recording as it features simultaneous sound sources at different distances.
Finally, METU-SPARG \cite{orhun_olgun_2019_2635758} features IRs recorded in an office, sampled over a 3D grid around the microphone. 
Distance information is present in its metadata files.
We use IRs collected below the microphone's center for testing, and the rest for training. 
Table \ref{datasets} summarizes datasets.
Because of the small number of recordings in LOCATA, MARCo, and METU-SPARG we use channel-swapping \cite{wang2023four} to augment the training set of these datasets by a factor of eight (not reflected in Table \ref{datasets}). 

\subsection{Model and loss}
We train a CRNN to dynamically estimate the distance of non-simultaneous sound sources.
We modify the CRNN published by Adavane et al. \cite{adavanne2018sound} to have two outputs: event detector $\hat{d}$ and distance estimator $\hat{y}$. 
$\hat{d}$ is trained with binary cross-entropy (BCE) and $\hat{y}$ with a regressor $\openE$. The model's loss is
\begin{equation}
    L = \frac{1}{N}\frac{1}{T}\sum_{n=0}^{N-1}\sum_{t=0}^{T-1}d_{n,t}\openE(y_{n,t},\hat{y}_{n,t}) + \text{BCE}(d_{n,t},\hat{d}_{n,t}),
\end{equation}
where $y \in \mathcal{R}^+$ and $\hat{y} \in \mathcal{R}^+$ are the true and estimated sound source distance, respectively. $d \in \{0,1\}$ and $\hat{d} \in [0,1]$ are the true and predicted sound presence, respectively. 
$N$ is the batch size, and $T$ is the corresponding model output length along the time dimension. 
$d_{n,t}$ multiplies $\openE$ to avoid distance estimates from contributing to the loss in the absence of sound events. 

Reducing the model's absolute or squared error prioritizes the accurate estimation of more distant sound sources. 
In other words, an error of 0.1 meters is more dramatic if the target is 1 meter away versus 10 meters away. 
Therefore, we also try using the absolute percent error and squared percent error, which should result in a loss that uniformly weighs errors across ground truth distances. 
Furthermore, we also experiment with the thresholded absolute percent error.
Table \ref{losses} shows the $\openE$ equations we experiment with.

\begin{table}
\begin{center}
\begin{tabular}{llllll}
\toprule\toprule
Model & Exp & Best $\openE$ & Mean $\downarrow$ & Median $\downarrow$ & Std $\downarrow$ \\ 
\midrule\midrule
    \multirow{4}{*}{CRNN} 
    
    & \multicolumn{1}{l}{TWL} & SPE &\multicolumn{1}{l}{0.413} & \multicolumn{1}{l}{0.330} & \multicolumn{1}{l}{0.347} \\\cline{2-6}

    & \multicolumn{1}{l}{TWA} & SE &\multicolumn{1}{l}{0.368} & \multicolumn{1}{l}{0.340} & \multicolumn{1}{l}{\textbf{0.244}} 
    
    \\\hhline{~=====}

    & \multicolumn{1}{l}{FWL-S} & APE & \multicolumn{1}{l}{\textbf{0.337}} & \multicolumn{1}{l}{0.290} & \multicolumn{1}{l}{0.246}\\\cline{2-6}

    & \multicolumn{1}{l}{FWL-D} & APE & \multicolumn{1}{l}{0.352} & \multicolumn{1}{l}{\textbf{0.269}} & \multicolumn{1}{l}{0.275}      

    \\\hline\hline
    avg pred    & & &0.452 & 0.410 & 0.283  \\ 
    \hline
    \cite{daniel2022echo} & & &0.448 & 0.326  & 0.416          \\ \hline
   \bottomrule
        \end{tabular}
    \setlength{\belowcaptionskip}{-15pt}
    \caption{Comparison of distance estimation performance on the LOCATA test set across experiments. For each experiment we report the model using the $\openE$ that resulted in the best cross-validation performance. in each experiment is also shown. TWL: train with LOCATA. TWA: train with all. FWL-S: fine-tune with LOCATA from STARSS, FWL-D: fine-tune with LOCATA from DCASE.}
    \label{tab:loc_results}
\end{center}
\end{table}

\begin{figure}[t]
  \centering
  \centerline{\includegraphics[width=\columnwidth]{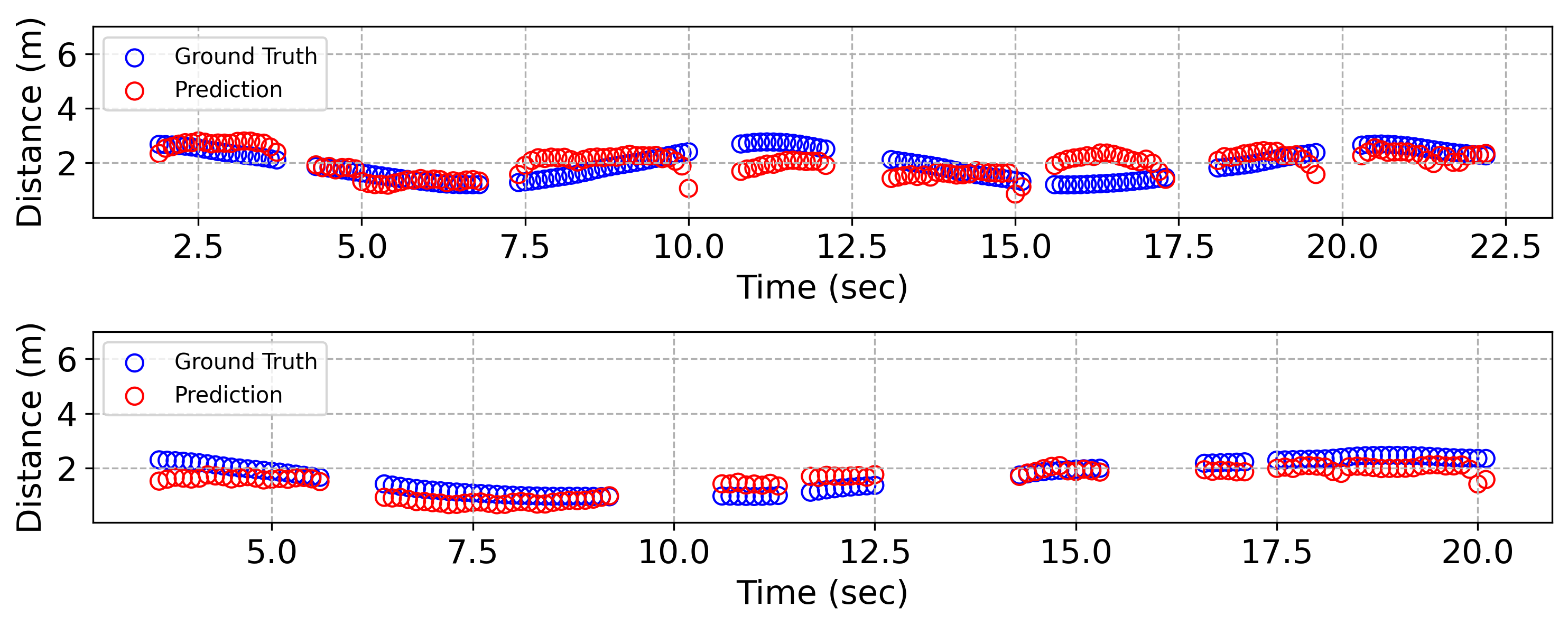}}
  \setlength{\belowcaptionskip}{-15pt}
  \caption{Qualitative comparison between the ground truth and best-model predictions on two excerpts from the LOCATA test set.}
  \label{fig:visualize}
\end{figure}

We keep the model's original input, consisting of a tetrahedral microphone's Log-mel spectrograms and generalized cross-correlation (GCC), capturing the difference in time of signal arrival between microphones.
To obtain the tretrahedral microphone (4 channels) from EigenMike (32 channels) recordings, we selected channels 6, 10, 26, and 22, consistent with the STARSS dataset \cite{politis_archontis_2023_7709052}.

\subsection{Training procedures}

First we trained the CRNN to carry out sound detection using the DCASE data.
During this phase the model's parameters are trained to optimally output $\hat{d}$, but the $\hat{y}$ output, the distance estimator, remains untrained.
We initialized model parameters with the Kaiming method and trained using an Adam optimizer (learning rate $1e-3$) with patience of 40 epochs based on optimal cross-validation performance (15\% of recordings randomly separated as a validation set).
This resulted on a sound event detector with an $F_1=0.94$ on the DCASE test set.
We refer to this model as the ``pre-trained sound event detector'' (PSED)\footnote{Pre-training $\hat{d}$ avoids local minima seen learning $\hat{d}$ and $\hat{y}$ from scratch.}.

Next, we initialized the CRNN with PSED parameters and we trained both $\hat{d}$ and $\hat{y}$ using LOCATA. 
We used Adam (learning rate $1e-3$) with patience of 40 based on optimal performance on a hold-out set consisting of the LOCATA ``Task 3 training'' files. 
We refer to this experiment as ``Train with LOCATA'' (TWL).
To study the potential benefit of using a larger training set, we repeated this experiment but substituted the training set to be all the training data across datasets listed in Table \ref{datasets}.
We refer to this experiment as ``Train with all data'' (TWA).

Compared to the DCASE and STARSS datasets, LOCATA is very small. 
Therefore, we also experimented with using LOCATA to fine-tune a model pre-trained with a larger dataset. 
We first initialized the CRNN with the PSED parameters to train both $\hat{d}$ and $\hat{y}$ using the STARSS dataset. 
We used Adam (learning rate $1e-3$) with patience of 40 based on optimal cross-validation performance on STARSS (15\% of recordings randomly separated as a validation set).
We refer to this as the ``STARSS pre-trained model'' (SPTM).
Next, we initialized the CRNN with the SPTM parameters to train both $\hat{d}$ and $\hat{y}$ using LOCATA. We used Adam (learning rate $1e-3$) with patience of 40 based on optimal performance on a hold-out set consisting of the LOCATA ``Task 3 training'' files. 
We refer to this experiment as ``Fine-tune with LOCATA from STARSS'' (FWL-S).
We also carried out this procedure using DCASE instead of STARSS, resulting in an experiment called ``Fine-tune with LOCATA from DCASE'' (FWL-D).
Each experiment is run seven times with a different regressor $\openE$: AE, SE, APE, SPE, TAPE($\delta=0.01$), TAPE($\delta=0.1$), and TAPE($\delta=0.2$), all listed in Table \ref{losses}.

\subsection{Baselines for comparison and metrics}

We compare performance on the LOCATA test set against the average sound source distance in the LOCATA training set (``avg pred''), and the recent signal processing approach by Daniel et al. \cite{daniel2022echo}.
It is worth noting that \cite{daniel2022echo} did not compare against a baseline since they consider their approach to be the first to not make assumptions about a room's DRR \cite{daniel2022echo}. 
In our own literature review, we did not find other distance estimation approaches evaluated on any of the open-source datasets that we consider in this study. 
We use the same metrics used in the study by Daniel et al. \cite{daniel2022echo}, which are the mean, median, and standard deviation of the model's absolute-valued distance estimate error.

\begin{table}
 \begin{center}
 \begin{tabular}{lllccc}
   \toprule\toprule
   Model & Exp & Best $\openE$&Mean $\downarrow$ & Median $\downarrow$ & Std $\downarrow$\\ 
   \midrule
    CRNN      & TW\underline{D}  & SE & 1.032 & 0.903 & 0.838 \\
    CRNN      & FW\underline{D}-S  & AE & \textbf{0.952} & \textbf{0.731}  & 0.834 \\
    avg pred  &   & &1.014 & 0.866 & \textbf{0.596} \\ 
    \hline
    CRNN      & TW\underline{M}  & SE & 1.346 & 0.417 & 2.158 \\
    CRNN       & FW\underline{M}-S & SPE & \textbf{0.811} & \textbf{0.405} & 0.508 \\
    avg pred  &  & &1.183 & 1.611 & \textbf{0.494}      \\ 
     \hline   
    CRNN      & TW\underline{T}  & APE  & \textbf{0.148} & 0.122 & \textbf{0.126} \\
    CRNN       & FW\underline{T}-S  & TAPE$^*$ & 0.167 & \textbf{0.114} & 0.150 \\ 
    avg pred  &   & &0.378 & 0.289 & 0.234 \\     
     \hline    
   \bottomrule
   \end{tabular}
\end{center}
\setlength{\belowcaptionskip}{-15pt}
 \caption{Performance on diverse datasets compared to the ``avg pred'' baseline. D=DCASE. M=MARCo. T=METU-SPARG. TWD and FWD-S highlight generalization to new rooms at test time. $^*$TAPE with threshold of $0.01$}
 \label{tab:all_datasets}
\end{table}

\section{Results}

Table \ref{tab:loc_results} shows that both ``avg pred'' and Daniel et al. \cite{daniel2022echo} baselines have similar ``mean'' metrics.
Thus it is possible that the method by Daniel et al. is correlated with the global statistics of the LOCATA training data \cite{daniel2022echo}.
Table \ref{tab:loc_results} also shows that all our experiments resulted in a model that outperforms both baselines.
The ``fine-tuning'' experiments (FWL-S and FWL-D) yielded the best performance according to the ``mean'' and ``median'' metrics.
To understand this pattern, let's remember that STARSS, DCASE and LOCATA consist of recordings in real rooms with humans producing sounds (i.e., speech, footsteps, etc.) around a microphone.
However, STARSS and DCASE have much more room diversity (16 rooms and 9 rooms, respectively) than LOCATA (1 room).
Therefore, initializing the CRNN with the SPTM model parameters (or the DCASE equivalent) may be providing with an initial representation of multi-room reverberation, from where it is easier to find the parameters to optimally perform in the acoustic conditions of the LOCATA room.
Figure \ref{fig:visualize} qualitatively compares predictions made by the best FWL-S model versus ground truth. 
Close alignment is observed, with errors still tracing the ground truth contour. 

The best models in Table \ref{tab:loc_results} may be overfitting. 
Contrasting ``fine-tuning'' experiments with TWL yields insight into this issue. 
TWL initializes the CRNN with PSED parameters, resulting in a distance estimator $\hat{d}$ that learns this task only on the LOCATA data.
This makes overfitting to the LOCATA training set likely and we do see poorer performance at test time. 
The better-performing TWA (based on the ``mean'' metric) shows the benefit of using more training data and significant mitigation of overfitting compared to TWL.

To further study this issue, we repeated the ``Train with \underline{LOCATA}'' and ``Fine-tune with \underline{LOCATA} from STARSS'' experiments with the other datasets:\underline{DCASE}, \underline{MARCo} and \underline{METU-} \underline{SPARG}.
Table \ref{tab:all_datasets} shows the results.
Compared to their ``avg pred'' baseline, we again see the benefit of initializing the CRNN with SPTM parameters vs PSED (on DCASE and MARCo according to the ``mean'' metric). 
However, this was not the case for METU-SPARG.
This can be explained by its small size and statistical properties that are virtually the same across training and test splits. 
Thus, overfitting to train data results in good performance on the test split. 

Tables \ref{tab:loc_results} and \ref{tab:all_datasets} also show what specific $\openE$ resulted in the best model.
In general, the ``percentage'' $\openE$s were better.
For FWL-S, we analyzed the effect of different $\openE$ (Table \ref{tab:losses}).
We observe that the ``percentage'' $\openE$s (APE, SPE, and TAPE) result in improved performance compared to AE and SE. 
This makes sense, as APE, SPE, and TAPE uniformly weight errors as a function of ground truth distance.
Figure \ref{fig:loss_compare} visualizes this effect by plotting the mean FWL-S model error as a function of ground truth distance for AE, APE, TAPE($\delta=0.01$), and TAPE($\delta=0.2$) on the LOCATA test-set.
Note also how AE tries to reduce errors associated with more distant sound sources and underperforms for sound sources that are closer to the microphone. 
In contrast, ``percentage'' $\openE$s  reduce prediction errors for targets closer to the microphone.
In general, performance deteriorates as a function of ground truth distance due to attenuation and arrival likely to be closely-followed by reverberations.

\section{Conclusion and future work}
We have proposed a model and optimization routine to carry out sound source distance estimation, which is an understudied component of SSL.  
Our solution is a CRNN with two outputs: a distance estimator and a sound event detector. 
Experiments revealed the benefit of using a loss function that uniformly weighs the model's estimate error across the full range of distances by converting it into a percentage of the ground truth distance. 
We also observe how the model tends to overfit to specific datasets, and the benefit of training with larger datasets featuring diverse acoustic conditions. 
To carry out this study, we have annotated sound source distances in a large collection of open-source datasets and a data generator, which we openly-release for future research by the broader community. 
In the future, we plan to expand this study by including more open-source datasets and adding more rooms to the data generator.
Future work could also investigate whether using the GTVV \cite{kitic2022generalized} as an additional or unique input feature to the model could improve performance. 
Similarly, other features like spectral magnitude difference \cite{georganti2013sound} and signal coherence \cite{vesa2007sound} or alternative input formats like larger microphone arrays, HOA or binaural audio could be used. 
Model architectures such as transformers and conformers could also be explored. 

Finally, a major shortcoming of the model presented here is its inability to track the distance of simultaneously-occurring sound sources. 
Recent solutions to this issue have been proposed in the DOA estimation literature \cite{shimada2022multi}, which could be applied to expand our approach.
Ultimately, we aim to develop a method that can jointly carry out the tasks of classification, localization, and distance estimation while being robust to different acoustic conditions.

\begin{table}
 \begin{center}
 \begin{tabular}{lccc}
   \toprule\toprule
   $\openE$ & Mean $\downarrow$ & Median $\downarrow$ & Std $\downarrow$ \\ 
   \midrule
    AE & 0.438 & 0.360 & 0.342 \\
    SE & 0.374 & 0.319 & 0.256 \\
    APE & 0.337 & 0.290 & 0.246  \\
    SPE & 0.334 & 0.292 & 0.259 \\
    TAPE ($\delta=0.01$) & 0.322 & 0.248 & 0.261 \\
    TAPE ($\delta=0.10$) & 0.361 & 0.312  &  0.250 \\
    TAPE ($\delta=0.20$) & 0.346 & 0.282 & 0.260 \\
     \hline    
   \bottomrule
   \end{tabular}
\end{center}
\setlength{\belowcaptionskip}{-10pt}
 \caption{The effect of different $\openE$ on the best-performing CRNN on the LOCATA test split. FWL-S experiments are shown since those yielded the best model. Note how TAPE($\delta=0.01$) has the overall best test-set performance. However, best model selection was agnostic of the test set, and was based on cross-validation performance to maximally emulate a real testing scenario.}
 \label{tab:losses}
\end{table}

\begin{figure}
  \centering
  \centerline{\includegraphics[width=\columnwidth]{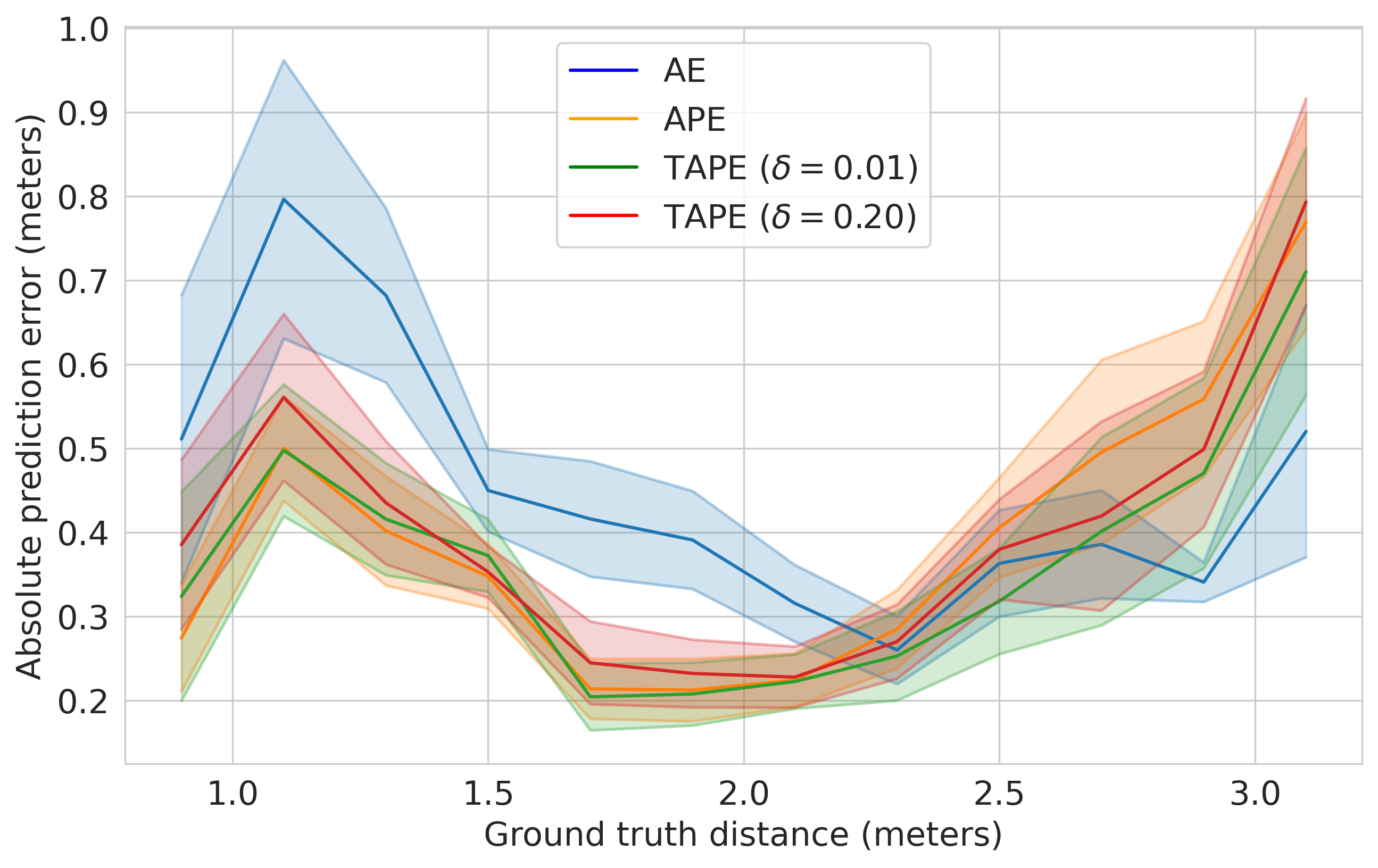}}
  \setlength{\belowcaptionskip}{-15pt}
  \caption{The effect of regressors $\openE$ on the CRNN distance estimate error as a function of ground truth distance (on the LOCATA test set in experiment FWL-S). Lines correspond to the CRNN's average error. The 95\% confidence interval of the mean is shown.}
  \label{fig:loss_compare}  
\end{figure}

\section{Acknowledgements}
This work is supported by the National Science Foundation grant no. IIS-1955357. The authors thank the funding source and their grant collaborators, particularly Bea Steers, who helped proof-reading this manuscript. 

\bibliographystyle{IEEEtran} 
\bibliography{refs23}

\end{sloppy}
\end{document}